\documentclass[10pt]{article}  
\usepackage[utf8]{inputenc}
\usepackage{amsmath}
\usepackage{amsfonts}
\usepackage{amssymb}
\usepackage{listings}
\usepackage{mathrsfs}
\usepackage{times}
\usepackage{float}
\usepackage{color}
\usepackage{setspace}
\usepackage{color,soul}

\usepackage{amsmath,amsfonts,amsthm,eucal}
\usepackage{enumerate}
\usepackage[letterpaper,margin=3cm]{geometry}
\usepackage{float}
\usepackage[title]{appendix}
\usepackage{color}
\usepackage{array}
\usepackage{comment}

\usepackage{multirow}
\usepackage{placeins}
\usepackage{bm}
\newcolumntype{M}[1]{>{\centering\arraybackslash}m{#1}}

\usepackage[
pdfstartview=XYZ,
bookmarks=true,
colorlinks=true,
linkcolor=blue,
urlcolor=blue,
citecolor=blue,
pdftex,
bookmarks=true,
linktocpage=true,   
hyperindex=true
]{hyperref}

\usepackage{lipsum}
\usepackage{lineno}
\usepackage[FIGBOTCAP,TABTOPCAP,bf,tight]{subfigure}

\usepackage{natbib}
\setcitestyle{authoryear}

\usepackage[pdftex]{graphicx}
\usepackage{epstopdf}
\usepackage{booktabs} 
\usepackage{lineno}
\usepackage{tikz,mathpazo}
\usetikzlibrary{shapes.geometric, arrows}
\usepackage{indentfirst}

\usepackage{caption}

\newcommand{\tensor}[1]{\ensuremath{\boldsymbol{#1}}}

\usepackage{algorithm}
\usepackage{algorithmicx}
\usepackage[noend]{algpseudocode}

\theoremstyle{remark}

\usepackage{algorithm}
\usepackage[noend]{algpseudocode}

\title{ A Large Language Model and Denoising Diffusion Framework for Targeted Design of Microstructures with Commands in Natural Language} 

\begin{document}


\author{
    Nikita Kartashov
    \and
    Nikolaos N. Vlassis\thanks{Corresponding author, Department of Mechanical and Aerospace Engineering, Rutgers University, Piscataway, NJ 08854. \textit{nick.vlassis@rutgers.edu}} 
}

\maketitle

\begin{abstract}
Microstructure plays a critical role in determining the macroscopic properties of materials, with applications spanning alloy design, MEMS devices, and tissue engineering, among many others. 
Computational frameworks have been developed to capture the complex relationship between microstructure and material behavior. 
However, despite these advancements, the steep learning curve associated with domain-specific knowledge and complex algorithms restricts the broader application of these tools. 
To lower this barrier, we propose a framework that integrates Natural Language Processing (NLP), Large Language Models (LLMs), and  Denoising Diffusion Probabilistic Models (DDPMs) to enable microstructure design using intuitive natural language commands. 
Our framework employs contextual data augmentation, driven by a pretrained LLM, to generate and expand a diverse dataset of microstructure descriptors. 
A retrained NER model extracts relevant microstructure descriptors from user-provided natural language inputs, which are then used by the DDPM to generate microstructures with targeted mechanical properties and topological features. 
The NLP and DDPM components of the framework are modular, allowing for separate training and validation, which ensures flexibility in adapting the framework to different datasets and use cases. 
A surrogate model system is employed to rank and filter generated samples based on their alignment with target properties.
Demonstrated on a database of nonlinear hyperelastic microstructures, this framework serves as a prototype for accessible inverse design of microstructures, starting from intuitive natural language commands. 
\end{abstract}

\section{Introduction}
\label{sec:intro}

Microstructure plays a critical role in determining macroscopic properties across various fields. Alloy phase distribution significantly affects mechanical properties \citep{wang1995aluminium,kang2019review}; microfeature precision in MEMS devices directly influences performance \citep{osiander2018mems}; and in tissue-engineered scaffolds, the cellular arrangement impacts cell growth and differentiation \citep{chen20153d}. 
The widespread relevance of microstructure in these and other domains is underscored by the search term "inverse design of microstructure," which yields approximately 546,000 results on Google Scholar as of July 2024, reflecting the vast scope of applications aimed at understanding the inverse mapping from material properties and behaviors to microstructure \citep{jain2014perspective,bostanabad2018computational}.

Numerous computational methods have been developed across fields to optimize microstructures for desired properties. 
Topology optimization, often paired with finite element methods (FEM), is used in multiscale mechanics and materials science to design efficient, high-performance components in industries like aerospace and automotive \citep{gao2000microstructure,xia2015multiscale}. 
Monte Carlo methods excel in exploring large configuration spaces, aiding in phase transitions, polymer dynamics, and crystallization in fields like metallurgy and pharmaceuticals \citep{meimaroglou2014review}. Genetic algorithms optimize multi-parameter systems in semiconductor design, biomaterials engineering, and other domains \citep{kim2005genetic, yeoman2009use}.

Generative artificial intelligence (AI) methods have significantly further advanced microstructure design. Variational autoencoders generate continuous microstructure spaces, enabling predictions of mechanical properties like stress-strain behavior using Gaussian process regression \citep{kim2021exploration}. Generative adversarial networks (GANs) are employed to synthesize complex materials with high spatial coherence, such as scaffold structures and metal foams \citep{zhang2021scaffoldgan}, and to optimize microstructures for specific property distributions in photovoltaic applications \citep{lee2021fast}. 
Combining GANs with reinforcement learning, synthetic microstructures with controlled physical properties have been generated for material discovery \citep{nguyen2022synthesizing}. Additionally, generative flow models create realistic 3D porous media structures, focusing on geosciences and petroleum engineering \citep{guan2021reconstructing}. 
Finally, denoising diffusion probabilistic models (DDPMs) have shown promise in synthesizing 3D sand grains with realistic morphology to design targeted granular assemblies in geotechnical applications \citep{vlassis2023synthesizing}.

Despite the considerable effort invested in these advancements, the broader potential of these tools remains untapped due to their limited cross-discipline applicability. 
The primary barrier to adopting these methods in new fields lies in the burden of knowledge, expertise, and time \citep{brewer1999challenges,burden,mazzocchi2019scientific}. These tools often come with specialized technical language, complex algorithmic structures, and extensive codebases, making them difficult to apply beyond their original scope. For an engineer or scientist seeking to employ these methods in a different domain, the time and expertise required to adapt them can be prohibitive. This burden creates a steep learning curve, and the economic and practical infeasibility of overcoming it slows innovation and problem-solving across disciplines \citep{van2020interdisciplinary}.

Meanwhile, Large Language Models (LLMs) \citep{chang2024survey} and Natural Language Processing (NLP) methods \citep{chowdhary2020natural} are gaining widespread adoption across various fields beyond computer science. In healthcare, NLP automates the analysis of clinical notes and predicts patient outcomes from unstructured data \citep{li2022neural}. In finance, it supports market analysis, fraud detection, and customer service automation \citep{gao2021review}. Legal fields use NLP for contract review and case law research, reducing time and expertise required \citep{hendrycks2021cuad}. In education, NLP tools enable automated grading, personalized learning, and feedback generation \citep{kasneci2023chatgpt}. These applications illustrate how NLP can simplify complex tasks across disciplines, making them more accessible through natural language commands.

In this work, we demonstrate how NLP methods and generative AI can be employed in mechanics to alleviate the knowledge burden in the inverse design of microstructures, enabling the design from natural language commands. Specifically, we introduce a framework that performs two key tasks: extracting microstructure descriptors from text and generating corresponding microstructures. The framework consists of an NLP component and a microstructure generation component. The NLP component leverages a large language model to create a text descriptor database and uses a retrained named entity recognition (NER) model to extract specific microstructure descriptors from natural language inputs. These extracted descriptors are then passed to the microstructure generation component, which employs a DDPM to generate microstructure samples with the desired properties. A system of surrogate models further refines the generated samples to ensure alignment with the specified properties. This approach reduces the expertise barrier and facilitates efficient microstructure design based on intuitive language inputs.

This paper is organized into the following sections. 
Section \ref{sec:why} will delve into the theoretical background and implementation of the components of the proposed framework. 
Section \ref{sec:database} describes the database of non-linear hyperelastic microstructures used to train the generative AI component of the framework and how it is expanded for NLP algorithms. 
Section \ref{sec:surrogate_models} introduces a system of surrogate models used to validate the generated microstructures.
Section~\ref{sec:num exp} details the training and testing results of the individual components of the proposed framework. 
Section \ref{sec:ensemble} demonstrates the deployment of the ensemble framework to generate microstructures starting from commands in natural language.  
Finally, concluding remarks and future directions are presented in Section~\ref{sec:conclusion}.

\section{Natural language processing and generative AI framework for microstructure generation from commands in natural language}
\label{sec:why}

\begin{figure}
    \centering
    \includegraphics[width=1\linewidth]{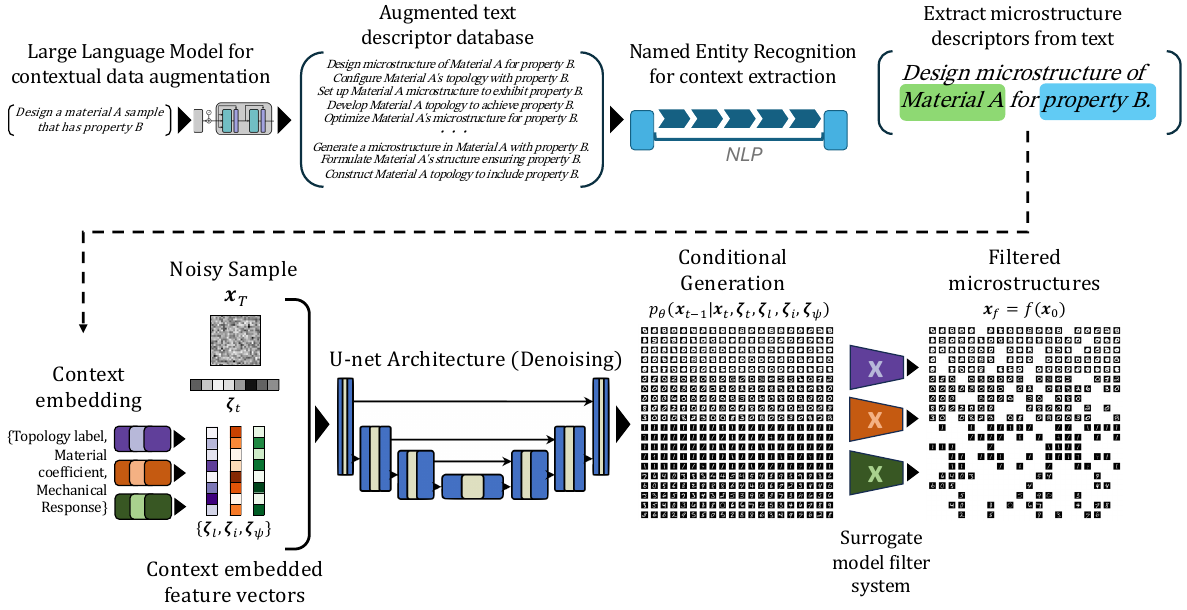}
    \caption{Schematic of overall framework components.
    The NLP component framework (top) involves a large language model for the generation and augmentation of text descriptor database and a retrained named entity recognition model to extract microstructure descriptors from text. 
    The microstructure generation component (bottom) involves a denoising diffusion probabilistic model framework and system of surrogate models for ranking and filtering the massively generate microstructure samples.}
    \label{fig:framework_schematic}
\end{figure}

The framework detailed in this paper integrates NLP techniques and generative modeling to streamline the design and generation of complex microstructures. 
The overall framework is divided into two main components: the NLP component and the microstructure generation component, as demonstrated in Fig.~\ref{fig:framework_schematic}. 
The NLP component utilizes an LLM to create and augment a text descriptor database to address the lack of a compiled text-based microstructure descriptor dataset in the literature. 
An NER model is then retrained on this database to extract specific microstructure descriptors from commands in natural language. 
The extracted descriptors are input to the microstructure generation component which employs a DDPM. 
This model generates microstructure samples with the extracted descriptor target properties by learning to reverse a conditional Markov diffusion process. 
Additionally, a system of surrogate models is implemented to rank and filter the generated microstructures, ensuring that the samples align with the desired properties specified in the input. 
This ensemble framework aims to lower the barrier to complex design, allowing for efficient microstructure generation starting from commands in natural language.

In Section~\ref{sec:llms}, the use of large language models for the generation and augmentation of text descriptor data sets for microstructures is introduced. In Section~\ref{sec:ner_algorithm}, we discuss our retrained NLP pipeline that extracts microstructure descriptors from text. In Section~\ref{sec:ddpm}, we describe the DDPM model framework that involves embedding the extracted microstructure descriptors in a latent space to perform conditional microstructure generation with targeted properties.

\subsection{Large language models for contextual data augmentation in mechanics}
\label{sec:llms}

Transformer-based architectures, initially proposed by \citep{vaswani2017attention}, revolutionized NLP -- the task of enabling computers to understand, interpret, and generate human language \citep{chowdhary2020natural}, by introducing the concept of self-attention mechanisms. 
This approach significantly improved the models' ability to capture long-range dependencies in text, surpassing the limitations of recurrent neural networks (RNNs) and convolutional neural networks (CNNs) \citep{yin2017comparative}. 
Subsequent developments on text embedding, such as the bidirectional 
encoder representations (BERT) model by \citep{devlin2018bert}, further refined the transformer architecture by incorporating pre-training techniques on large corpora, leading to state-of-the-art performance in a variety of NLP tasks.

The exponential growth in computational resources and the availability of generalist extensive annotated datasets have been pivotal in the development and refinement of LLMs \citep{narayanan2021efficient}. 
Models like GPT-3, trained on billions of tokens from the web, exemplify the scalability and generative capabilities of modern LLMs \citep{brown2020language}. 
This scale of training enables models to generate coherent and contextually relevant text, pushing the boundaries of what was previously achievable in automated language understanding and generation.
For more general areas of interests, large datasets are readily available through books, web texts articles, and other pieces of writing on the internet to train and fine-tune these models.
For more specific areas such as in finance \citep{wu2023bloomberggpt} and medicine \citep{thirunavukarasu2023large}, datasets have to be sourced, curated, and annotated carefully to allow for language understanding of domain-specific knowledge.

However, to our knowledge, there exists no obvious and well-curated database of text descriptors of microstructures, their properties, and their corresponding material responses.
While microstructure behavior description can be conveyed in natural language, it is heavily based on numerical \citep{anovitz2015characterization}, symbolic \citep{bahmani2024discovering}, or other data structure descriptors such as 2D and 3D images \citep{mohammed2018scanning,nguyen2022synthesizing}, series \citep{fullwood2008gradient}, level sets \citep{vlassis2021sobolev}, graphs \citep{vlassis2020geometric}, among many others.
Text descriptors are plentiful in the mechanics literature there is no compiled source of data that is readily available for retraining a domain-specific LLM.

In this work, we reconcile this gap in data by introducing a contextual data augmentation framework. 
Contextual data augmentation is a method that enhances labeled sentence datasets by replacing words within sentences with other contextually appropriate words often predicted by a bi-directional language model \citep{kobayashi2018contextual,wu2019conditional}. 
This technique operates under the assumption of invariance, which means that sentences remain natural and maintain their original labels even when words are substituted with others having paradigmatic relations. 
Unlike synonym-based augmentation, which is limited by the scarcity of exact synonyms, contextual augmentation uses a broader range of substitutes, ensuring a more diverse augmentation while preserving label compatibility. 
This method involves stochastic word replacements, guided by the surrounding context. Often quality control of this method is employed, such label-conditional architectures \citep{kobayashi2018contextual} to prevent replacements that might conflict with the sentence's original content, ensuring the generated sentences remain plausible and beneficial for model training.
Contextual data augmentation has been used to enhance NLP pipelines such as in machine translation \citep{gao2019soft} where available data has been sparse.

Contextual data augmentation is based on replacement of context-relevant words from a template sentences. While these templates can be readily available for general interest text, there exist none readily accessible and compiled in the mechanics literature.
We remedy this by borrowing a concept of information retrieval and text data augmentation using LLMs \citep{kumar2020data,feng2021survey} from the NLP literature.
LLMs have been previously used in augmenting text datasets to increase the volume and diversify data points by prompting \citep{yoo2021gpt3mix,bonifacio2022inpars,sahu2022data} or to generate data sets for domain-specific tasks where the data can be sparse or non-existent altogether \citep{tan2023inferring} usually paired with a retraining of language model or another NLP algorithm.
Similar to these works, we employ a pretrained transformer model, specifically the generative pretrained transformer GPT-4 \citep{achiam2023gpt} to massively generate template phrases to perform contextual data augmentation, eliminating user bias during data collection.

Thus, our microstructure text descriptor data generation and augmentation is two-fold.
First, we utilize a pretrained LLM to efficiently generate and curate template phrases from a single mechanics-specific prompt. The template prompts will hold placeholders of numerical and symbolic expression microstructure and behavior descriptors of the hyperelastic microstructures of this work. The details of the generation are discussed in Section~\ref{sec:data_augmentation}. Second, we performed contextual data augmentation based on these templates. The placeholder phrases will be replaced by real descriptors from a separate mechanics dataset of material property entries such as topology labels, material coefficients, and mechanical response expressions. This augmented dataset will be used to train and validate the NER model in this work as discussed in the following section.

\subsection{Named Entity Recognition}
\label{sec:ner_algorithm}

To incorporate natural language commands into the framework, text embedding was initially considered due to its current performance and utilization in established frameworks such as text-to-image generators \citep{rombach2022high, ramesh2021zero}. 
While these works succeeded in using embedding architectures to convey descriptions of general images, embedding architectures are not specialized to handle numerical data and other microstructure and mechanical response descriptors. 
Furthermore, embedding methods struggle with encoding information that includes critical alphanumeric data \citep{sundararaman-etal-2020-methods} where certain parts of the sentence, such as microstructure descriptors, are more important than the surrounding natural language commands.
This issue of precision in embeddings of semantically identical but different in formatting commands will also be discussed in Section \ref{sec:counterexample}.

To address these limitations,  NER was employed.
NER is a specific subtask of NLP that identifies and classifies key name entities within unstructured text, 
A named entity is defined as "a proper noun, serving as a name for something or someone", such as persons, locations, organizations, and geopolitcal enitities.
However, they are extended to concepts that do not fall into these categories such as dates, times, prices.
They also often involve multi-word phrases distinguishing NER from simple single word labeling (tagging).
More entity types can be reintroduced by retraining and fine-tuning these algorithms making them highly suitable for extracting relevant information from large volumes of natural language documents in various fields such as medicine \citep{10.1136/amiajnl-2011-000464} and finance \citep{zhao2021bert}.

Extracting these entities often involves a series of pre-processing steps to prepare the unstructured text which collectively ensure accurate identification and classification of the entities.
We base the implementation of our NER models on \cite{spacy2}.
This is a transition based approach instead of a traditional tagging framework implementation \citep{borthwick1999maximum,marquez2005semantic}. 
A transition based NER model is a finite-state automaton that operates on the sequential input of words as a queue derived from the input sentence similar to a stack data structure. 
The model then iteratively transitions between states using a trained convolutional neural network to select the optimal action from a predefined set. 
This action set included operations such as advancing through the queue, assigning tags to words, combining words into one entity, and several others. At each state the model uses the previous history of actions and decisions to determine which action is best for the current state, leading to the score of a state to be represented as used in \cite{watanabe2015transition} :
\begin{equation}
\rho(d) = \sum_{j=0}^{|d|-1} \rho(d_j|d_0^{j-1}),
\label{eq:NER State Score}
\end{equation}
where \(\rho\) is the score for a given decision \(d\), leading to the score of a decision to be dependent on all previous decisions made \(d_{0},d_{1},d_{2}...d^{j-1}_{0}\). And by using a neural models instead of linear trained models such as used in \cite{zhu2013fast}, a neural model can have a effectively infinite stack and queue history.

The implemented entity recognition architecture follows the "embed, encode, attend, predict" framework. 
Initially, each word in the text is transformed into dense embeddings, where key attributes such as the normalized form (lowercased version of the word), prefix, suffix, and word shape (pattern of letters and digits) are extracted. 
These attributes are then encoded though hashing, which involves mapping each attribute to a fixed-size embedding table using multiple random keys. 
This ensures that even unfamiliar words receive unique representations by spreading them across different positions in the table. 
These embeddings are concatenated and fed into a multi-layer perceptron (MLP), a type of neural network with multiple fully connected neural network layers, using maxout units to enhance learning capabilities, followed by layer normalization to stabilize and accelerate the training process, producing context-aware vectors for each word.

Across the algorithm, we come across two types of embeddings: non-contextual and contextual \citep{qiu2020pre}.
Non-contextual embeddings assign a single context-independent vector to each word, learned from a large corpus using self-supervision \citep{mikolov2013efficient}. 
However, they struggle with polysemy - words with multiple meanings are reduced to a single representation, and the out-of-vocabulary problem - they can only generate embeddings for words they have encountered in the training set. 
Contextual embeddings, on the other hand, provide token representations that consider the entire input sequence, capturing context-dependent word meanings. These embeddings are generated by neural contextual encoders \citep{li2020survey}, which use deep neural networks to encode context, achieving state-of-the-art performance on various NLP tasks when trained on large data.

So in this framework, the non-contextual encoded word vectors are further refined using CNNs, which capture contextual information from neighboring words. In this case, a trigram CNN layer is used to concatenate the vector representations of a token and its two neighbors. 
For example, a window of words on either side of the target word is concatenated to form a 384-dimensional vector. This vector is then processed through an MLP, reducing it back to 128 dimensions, effectively recalculating the word's meaning based on its context. 
This process is repeated across multiple layers, with residual connections by adding the input of each layer to its output and ensuring that the original information is preserved and preventing overfitting. 
Finally, a transition-based structured prediction framework is used, treating the NER task as a sequence of state transitions. The model starts with no output and iteratively processes each word, applying actions such as "begin entity" or "continue entity" to dynamically predict and classify named entities. 
This approach combines the strengths of tagging and parsing, ensuring flexibility and accuracy in identifying complex entity boundaries.

It is noted that we opt for this algorithm as it results in an efficient NER model that is computationally cheaper and faster than traditional recurrent neural networks  and the transformer models, while maintaining competitive performance \citep{li2020survey}.

\subsection{Denoising diffusion probabilistic models for conditional microstructure generation}
\label{sec:ddpm}

The framework for DDPMs originated from the work of \citet{sohl2015deep}, which introduced a class of generative models that learn to approximate a data distribution by reversing a multi-step, gradual noising process. 
Building on this, \citet{ho2020denoising} further advanced the concept, establishing DDPMs as equivalent to score-based generative models. These models learn the gradient of the log-likelihood of the data distribution via denoising score matching \citep{hyvarinen2005estimation}. 
DDPMs have demonstrated success in generating realistic data, with applications spanning image generation \citep{song2020denoising,ramesh2022hierarchical,rombach2022high}, audio synthesis \citep{chen2020wavegrad,kong2020diffwave}, and time-series forecasting \citep{rasul2021autoregressive}.
The DDPM architecture consists of two processes: a forward process and a reverse process. The forward process progressively applies Gaussian noise to the input data, corrupting the sample, while the reverse process, modeled using neural networks, learns to denoise the sample by approximating the reverse Gaussian noise process at each step. Training occurs using sample pairs at arbitrary timesteps during the noising process.

An important extension of DDPMs is conditional generation, where models are conditioned on specific inputs or descriptors to generate target outputs with desired characteristics. 
Conditional DDPMs have been successfully employed in generating microstructures based on predefined microstructural descriptors \citep{vlassis2023denoising}, allowing for targeted design in applications where specific material properties are required. 
This work reviews both the general DDPM framework and its adaptation for generating microstructures conditioned on such descriptors.

\subsubsection{Forward Process}
\label{sec:forward}
Provided a data set containing the distribution \(q(x_{0})\), a iterative forward noising process is introduced to process each sample \(x_{0}\) within the dataset to generate progressively noisy latent space data \(x_{1...T}\). This noising is defined as a Gaussian noise for timestep \(t\) with a variance schedule \(\beta_{t}\in(0,1)\), which is iteratively applied until time \(T\) where the original sample information is completely destroyed for very large $T$.  
The complete noising process is stated as the product of consecutive conditionals:
\begin{equation}
\label{eq:forwardTotal}
q(x_1, \dots, x_T | x_0) := \prod_{t=1}^T q(x_t | x_{t-1}),
\end{equation}
where the noising process between two timesteps is defined as: 
\begin{equation}
q(x_{t}|x_{t-1}) := \mathcal{N}(x_{t};\sqrt{1-\beta_{t}}x_{t-1},\beta_{t}I).
\label{eq:Markov Process}
\end{equation}

Using \(\alpha_{t} := 1-\beta_{t}\) along with \(\bar{\alpha}_t := \prod_{s=0}^t \alpha_s\), an arbitrary latent noisy sample at time \(t\) can be defined along with the forward step using
\begin{equation}
    q(x_t|x_0) = \mathcal{N}(x_t; \sqrt{\bar{\alpha}_t} x_0, (1 - \bar{\alpha}_t) \mathbf{I}),
\end{equation}
and the noisy sample would be
\begin{equation}
    x_t = \sqrt{\bar{\alpha}_t} x_0 + \sqrt{1 - \bar{\alpha}_t} \epsilon
\end{equation}
where \(\epsilon \sim \mathcal{N}(0, \mathbf{I})\) is a random variable.  Using the Bayes theorem, the posterior is found as
\begin{equation}
q(x_{t-1}|x_t, x_0) = \mathcal{N}(x_{t-1}; \tilde{\mu}_t(x_t, x_0), \tilde{\beta}_t I),
\label{eq:posterior}
\end{equation}
using the mean of the Gaussian and variance respectively as: 
\begin{equation}
\tilde{\mu}_t(x_t, x_0) := \frac{\sqrt{\bar{\alpha}_{t-1} \beta_t}}{1 - \bar{\alpha}_t} x_0 + \frac{\sqrt{\alpha_t (1 - \bar{\alpha}_{t-1})}}{1 - \bar{\alpha}_t} x_t  \text{ and } \tilde{\beta}_t := \frac{1 - \bar{\alpha}_{t-1}}{1 - \bar{\alpha}_t} \beta_t.
\label{eq:Gaussian mean variance}
\end{equation}

\subsubsection{Conditional reverse process for targeted microstructure generation}
The DDPM in this work is tasked to reverse the above noising process and perform conditional generation of microstructures -- that is sample microstructures given target properties. 
Conditional generation is achieved by expanding on \cite{vlassis2023denoising} which adapted the approach of \cite{ramesh2022hierarchical}.
Instead of generating images with textual descriptions and classes, we utilize embedded context vectors that carry the relevant material descriptor information for generating microstructure samples.

The generation is conditioned on $\tensor{y}_1, \tensor{y}_2,\dots,\tensor{y}_n \text { for }i = 1,2,\dots n$ material descriptors which will act as target input.
The descriptors hold information about the material topology, coefficients, and mechanical response.
For each desired target $\tensor{y}_i$,  a corresponding context embedding neural network $\tensor{\zeta}_i$ is defined,  embedding each target property into individual feature vectors $\tensor{\zeta}_i(\tensor{y}_i)$. 
The architecture of these embedding frameworks depends on the data structure representing the corresponding target input.
This allows for the flexibility and multi-modality of the framework --  multilayer perceptrons can be used for scalar and tensor values, convolutional neural networks for conditioning with images, graph encoders for use with graphs, and so on.

With individual embedded feature vectors generated, an embedded context vector is formed with:
\begin{equation}
\tensor{\zeta}_\text{context} = \tensor{\zeta}_1(\tensor{y}_1) + \tensor{\zeta}_2(\tensor{y}_2) + \dots + \tensor{\zeta}_n(\tensor{y}_n)  \text { for }i = 1,2,\dots n.
\label{eq:contextVector}
\end{equation}
Partial conditioning can also be performed where not all microstructure descriptors have to be provided.
For such cases, we input context $\tensor{y}_{i} = \tensor{y}_{i,\O}$, which is a pre-defined placeholder value in the corresponding data structure (e.g. a zero tensor).

The vector $\tensor{\zeta}_\text{context} $ can be representative of all possible combinations for context inputs $\tensor{y}_i$, and $\tensor{y}_{i,\O}$. 
For example, in our work we have \(n=3\) types of context inputs,  whose properties and their definitions are elaborated within section \ref{sec:database}. 
As such, $\tensor{\zeta}_\text{context} $ can contain all 8 possible combinations of inputs including unconditional cases ($\tensor{y}_{i} = \tensor{y}_{i,\O} \text{ for all } i$), conditional on one $\tensor{y}_i$, pairs of $\tensor{y}_i$, and fully conditioned cases with all three $\tensor{y}_i$ inputs provided. 
It is noted that these generated context vectors are invariant to the time vector and stay constant during the reverse process.

Provided the $\tensor{\zeta}_\text{context} $ vectors for each sample, the conditional reverse process $q(x_{t-1}|x_t, \tensor{\zeta}_\text{context})$ can be defined.
An approximation $p_{\theta}$ of the reverse process is learned by a neural network, usually a U-net \citep{ronneberger2015unetconvolutionalnetworksbiomedical}.
When $T\to \infty$ and $\beta_t \to 0$,  the distribution $q(x_{t-1}|x_t)$ to be learned approaches a diagonal Gaussian distribution.  
Thus,  the neural network can learn the mean $\mu_\theta$ and the diagonal covariance $\Sigma_\theta$, such that:
\begin{equation}
p_{\theta}(x_{t-1}|x_t,\tensor{\zeta}_\text{context} ) := \mathcal{N}(x_{t-1}; \mu_{\theta}(x_t, \tensor{\zeta}_\text{context},t), \Sigma_{\theta}(x_t, \tensor{\zeta}_\text{context},t)).
\label{eq:p_theta}
\end{equation}
\cite{ho2020denoising} demonstrate that the model can learn the mean of the process by predicting $\epsilon_\theta$ and we extend to generation conditioned on $\tensor{\zeta}$, such that:
\begin{equation}
\mu_\theta(x_t,\tensor{\zeta}, t) = \frac{1}{\sqrt{\alpha_t}} \left( x_t - \frac{\beta_t}{\sqrt{1 - \bar{\alpha}_t}} \epsilon_\theta (x_t,\tensor{\zeta}, t) \right).
\label{eq:learntmean}
\end{equation}
Following the improved formulation of DDPMs \citep{nichol2021improved}, our network also learns the variance of the process by estimating a vector \(v\) conditioned on $\tensor{\zeta}$, such that:
\begin{equation}
    \Sigma_\theta(x_t,\tensor{\zeta}, t) = \exp\left(v \log \beta_t + (1 - v) \log \tilde{\beta}_t\right).
\end{equation}
The model is optimized with the hybrid learning objective:
\begin{equation}
L_{\text{hybrid}} = L_\mu + \lambda L_{\text{vlb}},
\label{eq:hybrid Objective}
\end{equation}
where $\lambda=0.001$ is a scaling factor of the loss terms $L_\mu$ and $L_{\text{vlb}}$ and set to $0.001$ as used by \cite{nichol2021improved}. The term $L_\mu$ optimizes the approximated $\mu_\theta$, such that:
\begin{equation}
L_\mu = \mathbb{E}_{t, x_0, \epsilon} \left[ \| \epsilon - \epsilon_\theta(x_t, \tensor{\zeta},t) \|^2 \right],
\label{eq:MSE}
\end{equation}
by randomly sampling $t$ to estimate a variational lower bound.
The term $ L_{\text{vlb}}$ optimizes the approximated $\Sigma_\theta$ using:
\begin{equation}
\label{eq:vlb}
    L_{\text{vlb}} := L_0 + L_1 + ... + L_{T-1} + L_{T},
\end{equation}
where:
\begin{equation}
L_t := \begin{cases}
   - \log p_\theta(x_0|x_1) & \text{if } t = 0, \\
   D_{KL}(q(x_{t-1}|x_t, x_0) || p_\theta(x_{t-1}|x_t)) & \text{if } 0 < t < T, \\
   D_{KL}(q(x_T|x_0)||p(x_T)) & \text{if } t = T.
   \end{cases}
\label{eq:Lvlb}
\end{equation}
Each term in the above optimizes the Kullback-Leibler (KL) divergence between the Gaussian distributions of two successive denoising steps:
\begin{equation}
D_{K L}\left(p_\theta \| q\right)=\int_x p_\theta(x) \log \frac{p_\theta(x)}{q(x)}.
\end{equation}

It is noted that the weights of the embedding networks $\tensor{\zeta}_i$ will be optimized with back-propagation along the DDPM U-Net model using the same loss function in Eq.~\eqref{eq:hybrid Objective}. 

In this work, the reverse process is approximated by a U-net architecture.
It is a modified version of the implementation of \cite{nichol2021improved}, that was based on the image datasets ImageNet \citep{deng2009imagenet} and CIFAR-10 \citep{krizhevsky2009learning} of size 64 by 64 pixels along with the typical 3 channels for color. 
The property distribution maps used within this work are 28 by 28 pixels in size and have a single channel, as such a reduced size of the U-net architecture was used. 
The configured U-net contains 4 total steps comprised of 2 down-sampling and 2 up-sampling steps, each step containing 3 residual blocks. 
Eight attention heads \citep{vaswani2017attention} with identical amounts of channels were used. 
The base modal channel size was selected as $C=64$, with the U-net channel depth pattern following $[C,2C,4C]$. 
For the time vector used in the forward and reverse process, a sinusoidal positional embedding for time was implemented to compute unique representations of each time step in a continuous domain.

\FloatBarrier
\section{Training Database}
\label{sec:database}
In this section, we describe the database used to train and validate the models in this work. In Section~\ref{sec:mnist_descriptors}, we introduce the database of microstructures and the descriptors we use to describe them. In Section~\ref{sec:data_augmentation}, we discuss how we perform the contextual data augmentation to acquire text-based descriptors to train the NLP components of our framework.

\subsection{Mechanical MNIST with additional descriptors}
\label{sec:mnist_descriptors}

The training database that this work is based on is an extension of the MNIST dataset of handwritten digits \citep{deng2012mnist} -- a classical computer vision benchmark dataset. 
We utilize the extension of this dataset, Mechanical MNIST \citep{lejeune2020mechanical}, a database that treats the 28 by 28 grayscale images of the original database as maps of hyperelastic material property distributions. Specifically, every image is a distribution of a Young's modulus $E$ following:
\begin{equation}
E=\frac{\beta}{255}*(100-1)+1,
\label{eq:youngs_mod}
\end{equation}
where $\beta$ is the grayscale pixel intensity.
The higher the pixel intensity, the stiffer the material at that point. 
The digit corresponds to a stiff inclusion in a soft matrix, with the pixels with a maximum intensity corresponding to a point with two orders of Young's modulus higher than the soft matrix.
The Poisson ratio is considered constant, $\nu= 0.3$. The underlying model is a Neo-Hookean energy functional.
\begin{figure}
    \centering
    \includegraphics[width=0.95\linewidth]{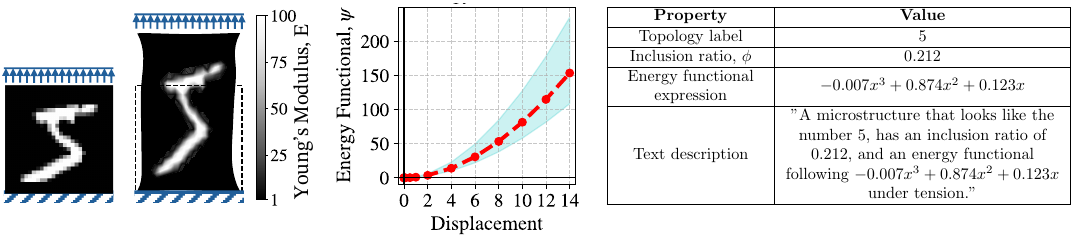}
    \caption{A sample of the training dataset of the framework consists of the FEM recorded homogenized energy functional under tension as well as the topology label, the inclusion ratio, a symbolic expression of the energy functional, and a text description of the microstructure.}
    \label{fig:sample}
\end{figure}
The Mechanical MNIST dataset consists of 60,000 training and 10,000 testing pairs of property distribution maps, and finite element simulations including uniaxial tension, shear, biaxial tension simulations among others. We use the uniaxial extension Test subset for this study. 
From every finite element simulation, we utilize the homogenized energy functional curve, $\psi$, over 13 imposed displacement steps ($d = \left[0.0, 0.001, 0.01, 0.1, 0.5, 1.0, 2.0, 4.0, 6.0, 8.0, 10.0, 12.0, 14.0 \right]$) as demonstrated in Fig.~\ref{fig:sample}. 
We expand this database with additional microstructure and material response descriptors to benchmark our NLP and microstructure generation tasks. 

In this work, every sample is characterized by a topology label describing the shape of the stiff inclusion that corresponds to the original MNIST digits, integers from 0 to 9.
We also introduce a metric of the size of the inclusion. The scalar inclusion ratio is defined as: 
\begin{equation}
\phi = \frac{A_\text{stiff}}{A_\text{total}},
\label{eq:inclusion_ratio}
\end{equation}
where $A_\text{stiff}$ is the area of the stiff inclusion, and $A_\text{total}$ is the total area of the microstructure. The points, or pixels, that are counted in the stiff inclusion have a Young's modulus larger than 1. The Young's modulus of the soft matrix is 1, corresponding to a pixel intensity of 0.

To facilitate the introduction of the material response in our text-based algorithms, we represent the recorded energy functional $\psi$ from the FEM simulations as a symbolic expression. We perform symbolic regression on the 70,000 $\psi$ curves of the training and testing datasets to acquire polynomial expressions of the form $\alpha x^3 +\beta x^2 +\gamma x$ for every sample. The symbolic expressions are stored as string variables. The original $\psi$ curve can be recovered by evaluating the symbolic expression. It is noted that the polynomial regression may lead to small loss in precision of the recorded psi representation. However, it is considered minimal in this work, with the main focus being a simple textual representation. More complex and precise symbolic representations will be considered in future work.

The last descriptor introduced in this work is a text prompt description of a microstructure. Stored as a string variable, the text prompts carry information about the microstructure, the shape and size, as well as its hyperelastic response, including all or a subset of the topology label, inclusion ratio, and $\psi$ symbolic expression. Construction of these prompts will be discussed in the following section.

Thus, every sample in the dataset is described by a Young's modulus distribution map, a set of 13 energy functional recorded values, an integer topology label, a scalar inclusion ratio, a symbolic expression of $\psi$, and a text prompt. An example of these descriptors can be seen in Figure~\ref{fig:sample}.

\subsection{Mechanics contextual data augmentation with LLMs}
\label{sec:data_augmentation}

The main challenge of training our NLP framework component was the lack of a text-based dataset of microstructure descriptors. 
Manually compiling sets of phrases was an option. 
However, the effort required and the risk of high bias in the dataset from using a small number of writers led to the decision to opt for a LLM-based data augmentation.

The database of the text prompt descriptors introduced in this work is acquired through contextual data augmentation as described in Section~\ref{sec:llms}.
To perform the data augmentation a set of template prompts and a set of context descriptors are required.
The template prompts are acquired through the paraphrasing of a master template with the GPT-4 architecture \citep{achiam2023gpt}. As a master prompt, we utilize the phrase: "Generate a microstructure that looks like the number MNIST\_ LABEL, has an inclusion ratio of INCLUSION\_ RATIO, and an energy functional of ENERGY\_ FUNCTIONAL." 
MNIST\_ LABEL, INCLUSION\_ RATIO, and ENERGY\_ FUNCTIONAL are placeholders for which the data augmentation is performed. 
GPT-4 is tasked to paraphrase the master prompt continuously until 150 prompt templates are collected. 
Duplicate prompts are automatically removed and prompts for which the paraphrasing is considered unsuccessful are manually deleted. 
The complete set of 150 templates is evaluated and split into sets of 100 and 50 for generating training and testing sets respectively.

These sets are then used as the basis for contextual data augmentation. We randomly subsample templates from the training and testing sets of the generated template pool. We also randomly select topology labels, inclusion ratios, and symbolic expression descriptors from the training and testing sets of the Mechanical MNIST database as described in the previous section. The placeholders in the prompts are substituted with the selected descriptors to generate text descriptors to train and validate our NER model.
Specifically, we generate 5,000 training and 10,000 testing prompts with replacement.
It is noted that when selecting the descriptors to replace in the templates, we do not necessarily select descriptors from the same mechanical MNIST sample. The success of the training of the NER model will be evaluated for the recognition of the descriptors for the given prompt syntax, and will be unaffected by the relationship between descriptors.

\section{Surrogate model filter system} 
\label{sec:surrogate_models}

The probabilistic nature of the microstructure generation in this work may lead to unexpected properties in the samples. To rank and filter out the generated microstructures that do not fulfill the expected property standards, we introduce a filter system that consists of offline trained neural networks and post-hoc calculations. Specifically, we train neural network topology label classifier, a neural network energy functional predictor, and post hoc calculate the generated inclusion ratios.

To validate the stiff inclusion shapes, we train a single classifier neural network architecture to identify the topology labels in the generated samples.
The neural network implemented in this script is a simple classifier designed using a fully-connected architecture. It consists of three linear layers: the first layer takes the flattened 28x28 pixel images from the MNIST dataset and outputs 1024 features. The second layer reduces the 1024 features to 512, and the final layer outputs 10 features corresponding to the 10 classes of the MNIST dataset. The ReLU activation function is applied after the first two layers, and the log-softmax function is used at the output layer to produce probability distributions over the classes.
Preprocessing involves normalizing the topologies using standard mean and standard deviation values.
The model is trained using the Negative Log-Likelihood loss function and the NAdam optimizer with a learning rate of 0.01. Training is conducted for 100 epochs with a batch size of 64. A scheduler adjusts the learning rate by a factor of 0.9 if the validation loss plateaus for 2 epochs. The model was trained on the 60,000 and tested on the 10,000 MNIST sets.
The model achieves an average loss of 0.1685 and an accuracy of 96\% for the training set
and an average loss of 0.2126 and an accuracy of 95\% for the testing set.
It is noted that a more robust classifier could be trained but as it will be discussed later, the classification test in this work was performed as a sanity check and in small generation batches the labels can be manually checked individually.

The second surrogate neural network implemented follows the same architecture as described in \cite{vlassis2023denoising} and is intended to replace costly full-scale finite element uniaxial extension simulations to validate the generated microstructures demonstrating the target energy functionals. 
The model inputs the Young's modulus distribution maps and outputs a hyperelastic energy functional curve.
The model consists of two convolutional layers, each with a kernel size of 3, stride of 1, and padding of 1. The first convolutional layer has 1 input channel and 16 output channels, while the second layer has 16 input and output channels. The output of the first convolutional layer is down-sampled by a Max Pooling layer with a kernel size of 2 and stride of 2, which selects the maximum value over a 2x2 window. The second convolutional layer's output undergoes the same down-sampling process. This output is then flattened into a single dimension to represent the extracted features. Subsequently, the flattened output is fed into three fully-connected layers with 120, 84, and 13 neurons, respectively. The output size of 13 corresponds to the number of time steps in the Mechanical MNIST uniaxial extension curves. ReLU activation functions are applied to the outputs of the CNN and dense hidden layers, while the final output uses a Linear activation function. The kernel weight matrices are initialized with the Glorot uniform distribution, and the bias vectors are initialized with zeros. The model was trained for 200 epochs with a batch size of 256 using the Adam optimization algorithm at a learning rate of $10^{-3}$ and a mean squared error loss function. The learning rate was reduced by a factor of 0.9 when the validation loss plateaued for 5 epochs. Training was conducted on 60,000 MNIST samples, with validation on an additional 10,000 samples, including their corresponding energy functional curves. The model achieved an average MSE of $3.96\times10^{-4}$ and $4.04\times10^{-4}$ for the training and testing sets of the Mechanical MNIST database.

The final filter in the system is the inclusion ratio validation for the generated samples which does not require a surrogate model as it can quickly post hoc calculated with Eq.~\ref{eq:inclusion_ratio}. 
By setting hyperparameter cut-off MSEs, we can quickly filter out the generated samples that do not adhere to the target inclusion ratio and energy functionals.
The surrogate model filter system will be employed to efficiently validate the massively generated samples in our verification experiments and the deployment of our ensemble framework.
It is noted that surrogate models are opted for due to the simplicity and speed of implementation. All these checks, can be performed manually for the topology label classification or more robustly with FEM simulations for the energy functionals. However, they were considered adequately accurate for the scope of this work.

\section{Numerical experiments} 
\label{sec:num exp}

In this section, we demonstrate the training and validation results for the proposed framework components.
In section \ref{sec:counterexample}, we investigate text embedding in the context of generation of microstructures and reason why we opt for an NER approach instead.
In section \ref{sec:nerresults},  we discuss the training and testing performance of the NER implementation in recognizing the novel mechanics-related named entities within the natural language prompts. 
In section \ref{sec:ddpmValidation},  we validate the DDPM algorithm though a comprehensive list of tests showcasing its performance in generating microstructures with targeted properties using combinations of prompts of increasing difficulty. 
In section \ref{sec:filtering},  we demonstrate the generated sample filtering process that ranks generated microstructures to adhere to target design constraints.

\subsection{Embedding counterexample} 
\label{sec:counterexample}
In this work, we aim to generate microstructures with specific properties based on textual descriptions. 
As discussed in Section~\ref{sec:ner_algorithm}, embedding text descriptions into a high-dimensional space is often used to process and utilize semantic information contained in the text effectively. However, this embedding process can introduce issues if not handled carefully, particularly when dealing with numerical data.
In this section, we provide a counterexample for why we opt not to use text embedding for our microstructure descriptors.

For instance, consider the sentences "Generate a microstructure that has an inclusion ratio of 0.3" and "Generate a microstructure that has an inclusion ratio of 0.30." Despite being semantically identical, these sentences may produce different embeddings due to the slight variation in numerical formatting. 
This discrepancy arises because models like BERT \citep{devlin2018bert}, which is used for natural language processing tasks, tokenize and process text at the subword level. 
Even minor variations in numerical formatting can lead to different token sequences, resulting in distinct token representations and ultimately different embeddings. BERT generates contextual embeddings for each token by considering both left and right context in all layers of its algorithm.

\begin{figure}
    \centering
    \includegraphics[width=.8\linewidth]{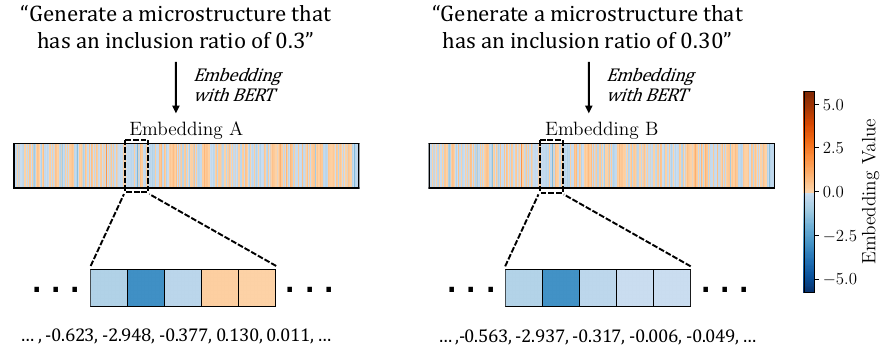}
    \caption{Embedding of two semantically identical microstructure descriptions with different formatting of descriptors leads to different text embeddings.}
    \label{fig:counterexample}
\end{figure}

To illustrate this issue, we generated embeddings for the two sentences using BERT. 
BERT provides 768-dimensional embeddings that encapsulate the semantic meaning of the input text by transforming the input tokens through multiple layers of attention mechanisms. Despite the overall similarity between the embeddings, with a high cosine similarity of 0.993, they are not identical. This slight variation in the embeddings, as depicted in Fig.~\ref{fig:counterexample}, occurs because BERT processes the text at the subword level, meaning that even a minor difference in numerical formatting (e.g., "0.3" vs. "0.30") can result in different token sequences and hence different embeddings.
The cosine similarity score indicates that while the two embeddings are close in the high-dimensional space, the minor discrepancies can still impact downstream applications. 
In the context of our work, this small difference in embeddings can lead to variations in the generated microstructures and, most importantly, loss of precision as the generative model might interpret identical prompts differently. 
This demonstrates the sensitivity of neural language models to input formatting, underscoring the need for consistent representation of numerical data to ensure reliable and accurate generation of microstructures that meet the intended specifications.

In this paper, we address this issue by employing an NER algorithm to standardize generation targets, ensuring that they produce consistent embeddings as described in the following section. 
By extracting and normalizing numerical data and other key entities, we can mitigate the impact of formatting differences, leading to more reliable and consistent prompts.

\FloatBarrier

\subsection{NER Training and Testing}
\label{sec:nerresults}

\begin{figure}
\centering
\includegraphics[width=.9\linewidth]{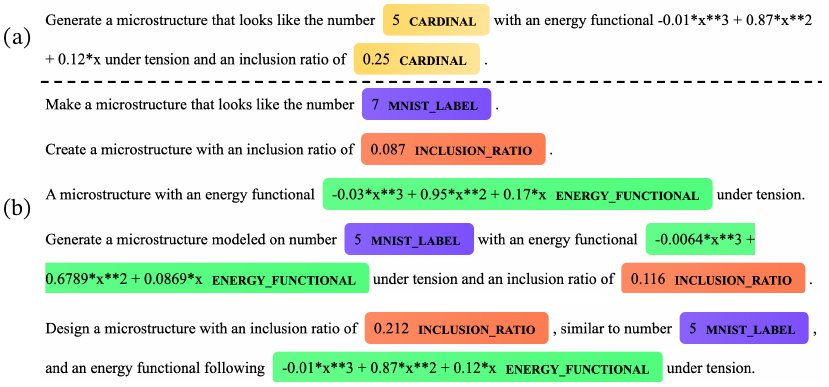}
\caption{(a) Results for one prompt before retraining the NER model. (b) NER results for 5 prompts and contexts from the Mechanical MNIST testing set after retraining.}
\label{fig:ner_results}
\end{figure}

In this section, we demonstrate the results of our retrained NER model described in Section ~\ref{sec:ner_algorithm}. Specifically, we utilized 5000 context-augmented prompts derived from 100 template prompts and the Mechanical MNIST training set, as detailed in Section~\ref{sec:data_augmentation}. 
For testing, we used 10,000 context-augmented prompts from 50 template prompts and the Mechanical MNIST testing set, ensuring these prompts were not seen during training. 
As the basis pre-trained pipeline model, we used the "en\_core\_web\_lg" model, which is a large English NLP pipeline for tasks such as tokenization, part-of-speech tagging, dependency parsing, and named entity recognition that can recognize and process 514,157 different words or tokens -- each word or token is mapped to a 300-dimensional vector space for semantic representation.

To retrain the NER model, we first compiled a training dataset comprising text descriptions paired with annotated entities. These entities included the MNIST label, inclusion ratio, and energy functional symbolic expression. The training process involved iteratively updating the model's parameters to optimize its ability to accurately recognize and classify these entities within the given text. The dataset was preprocessed to standardize the entity annotations, ensuring consistent formatting and alignment with the target labels. The annotated entities were formatted by specifying their location in the text using start and end indices, along with the entity type name for each data sample.

During the training process, we froze the weights for all models in the pipeline except for the NER component. 
We introduced three new entity labels into the NER model: "INCLUSION\_RATIO", "MNIST\_LABEL", and "ENERGY\_FUNCTIONAL". The training was conducted by iterating through the annotated examples for 25 epochs. We used the Stochastic Gradient Descent (SGD) optimizer with a learning rate of 0.001.

We evaluated the accuracy of the retrained NER model in extracting the MNIST label, inclusion ratio, and energy functional symbolic expression. Successful recognition was defined as the entire string being correctly identified and matching the target string exactly. The retrained NER model achieved accuracy scores of 100\% for MNIST labels, 100\% for inclusion ratios, and 94.07\% for energy functional symbolic expressions. The errors in recognizing energy functional symbolic expressions typically involved missing terms or failing to fully recognize all polynomial terms.

As shown in Fig.~\ref{fig:ner_results}, the results highlight the model's capability in accurately extracting the specified entities from sample prompts and contexts from the Mechanical MNIST testing set. Given a test prompt before training (Fig.~\ref{fig:ner_results}~(a)), the basis model can only identify that there exist numerical values in the prompt (labeled as "CARDINAL") but cannot differentiate between inclusion ratios and topology labels. After training (Fig.~\ref{fig:ner_results}~(b)), the model has successfully introduced and can recognize the three new entity types. 

It is noted that the identified entities will be further evaluated before being input to the DDPM model, with a step of visualization for user verification (Section~\ref{sec:ensemble}). 
In cases where an entity is not identified correctly, a warning is issued to the user.
In future work, we plan to consider using a transformer-based NER algorithm or fine-tuning a language model to enhance performance. However, for this study, we opted for a simpler NER model for ease of implementation, training, and deployment. 

\FloatBarrier

\subsection{Validation of the DDPM architecture} 
\label{sec:ddpmValidation}
To validate the modularity of the DDPM, we devised a comprehensive testing matrix and evaluated the trained model against it. 
This matrix consists of conditional generation tasks of increasing complexity, encompassing combinations of MNIST labels, inclusion ratios, and energy functionals. 
Our aim was to test the model's ability to generate microstructures that meet specific criteria while maintaining high performance across different conditional settings.

We first validated whether the model performs well across any combination of the learned prompt contexts. It is noted that this validation step focuses solely on the internal representations learned by the model and is completely decoupled from the natural language text prompts. By testing these combinations, we ensure that the DDPM can effectively handle and generate microstructures based on different combinations of contextual information -- crucial for verifying the model's capabilities before integrating it with natural language commands in the following section. 
\begin{table}[h!]
    \centering
    \begin{tabular}{|c|c|c|c|}
        \hline
        \text{Test ID} & \text{Code} & \text{Context} & \text{Number of Prompts} \\
        \hline
        1 & \text{LL} & \text{MNIST label} & 10 \\
        \hline
        2 & \text{II} & \text{Inclusion Ratio} & 4 \\
        \hline
        3 & \text{EE} & \text{Energy Functional} & 3 \\
        \hline
        4 & \text{LI} & \text{MNIST label, Inclusion Ratio} & 40 \\
        \hline
        5 & \text{LE} & \text{MNIST label, Energy Functional} & 30 \\
        \hline
        6 & \text{EI} & \text{Inclusion Ratio, Energy Functional} & 12 \\
        \hline
        7 & \text{LEI} & \text{MNIST label, Inclusion Ratio, Energy Functional} & 120 \\
        \hline
    \end{tabular}
    \caption{Description of tests and corresponding contexts for the evaluation of the trained DDPM architecture.}
    \label{tab:tests_contexts}
\end{table}

The testing matrix is structured to cover a broad range of conditional contexts, organized into seven distinct tests, each with varying levels of complexity and combinations of target properties. The detailed structure of the testing matrix is show in Table~\ref{tab:tests_contexts}.
This matrix was designed to systematically challenge the DDPM with a variety of prompts that require generating microstructures under different contextual conditions.
In constructing this matrix, we combined the ten MNIST labels with four target inclusion ratios, specifically $\{0.1, 0.2, 0.3, 0.4\}$, and three recorded energy functionals under tension from the mechanical MNIST testing set: one with low stored energy at the final displacement, one with energy close to the average, and one with high stored energy. 
These combinations were chosen to ensure that the DDPM was thoroughly tested across a diverse set of conditions, spanning all the possible combinations of the training data ranges. 

Test 1 (LL) focuses on generating microstructures based solely on MNIST labels, providing a baseline for the model's ability to recognize and generate specific digit shapes. 
Test 2 (II) and Test 3 (EE) individually assess the model's performance in generating microstructures with targeted inclusion ratios and energy functionals, respectively.
Tests 4 (LI), 5 (LE), and 6 (EI) increase the complexity by combining two contextual factors. 
Test 4 (LI) involves generating microstructures conditioned on both MNIST labels and inclusion ratios, while Test 5 (LE) combines MNIST labels with energy functionals, and Test 6 (EI) tests the combination of inclusion ratios and energy functionals.
The most complex test, Test 7 (LEI), involves generating microstructures based on all three contextual factors: MNIST labels, inclusion ratios, and energy functionals. This test comprises 120 unique prompts and represents the highest level of challenge for the DDPM, assessing its ability to integrate multiple contextual inputs simultaneously.
In total, we have 219 unique prompts, and for each prompt, we generate 100 samples, resulting in 21,900 samples for which we evaluate the performance of the DDPM.

\begin{figure}
    \centering
    \includegraphics[width=1\linewidth]{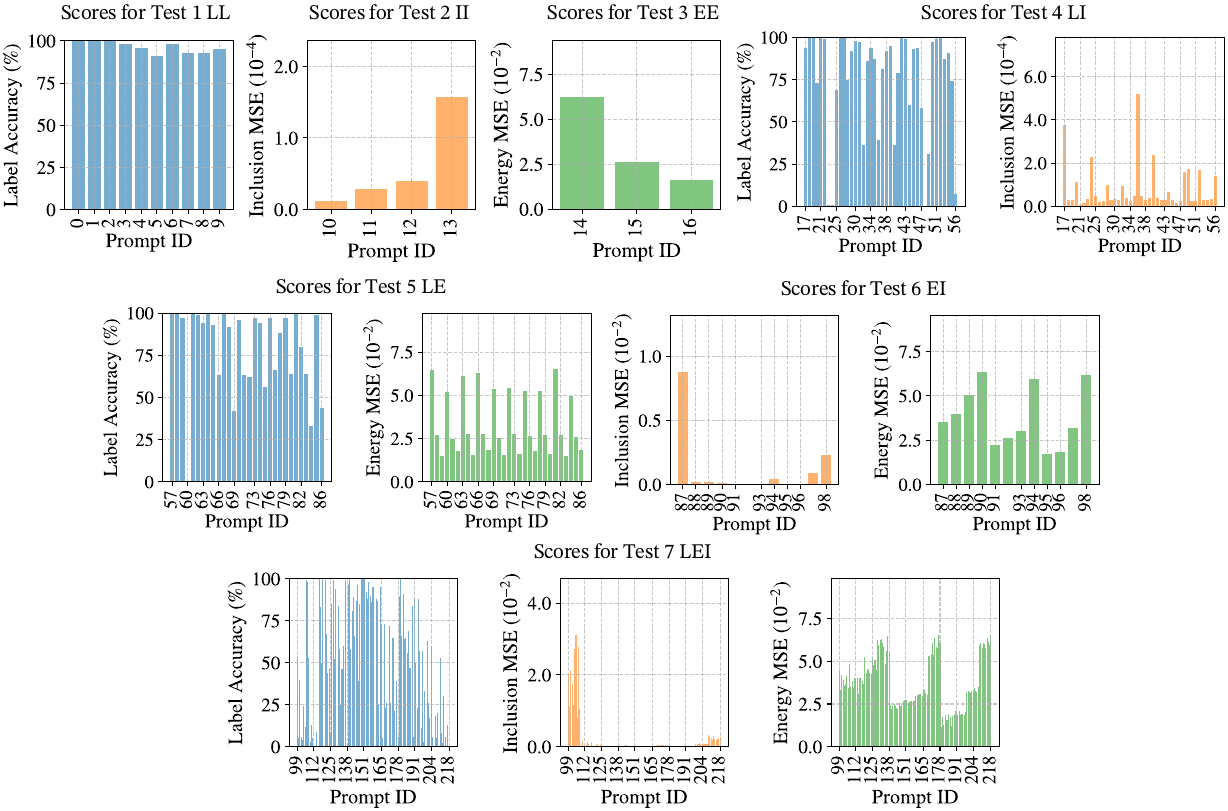}
    \caption{Overall performance of all the 219 combination prompts for the seven tests described in Table~\ref{tab:tests_contexts}.}
    \label{fig:all_scores}
\end{figure}

We evaluate the generated samples' performance using the surrogate model system as described in Section~\ref{sec:surrogate_models}. 
Depending on the type of test, we check for the accuracy of the MNIST label predicted with the trained classifier, 
post-process the inclusion ratio of the generated sample using Eq.~\eqref{eq:inclusion_ratio} and calculate a mean squared error with the target prompt, and predict the energy functional using the trained CNN surrogate model, then calculate the average MSE with the target energy prompt (if such a comparison is valid for the test type). 
For each of the 219 prompts, we calculate the average scores over the 100 generated samples.
The overall performance for these metrics for each of the prompts of the 7 tests are shown in Fig.~\ref{fig:all_scores}. 
As expected, the overall error distributions show that simpler prompts generally achieve lower errors, while more complex prompts exhibit higher error rates. Some prompts perform significantly better than others, highlighting the model's varying effectiveness depending on the complexity of the generation task.

\begin{figure}
    \centering
    \includegraphics[width=.85\linewidth]{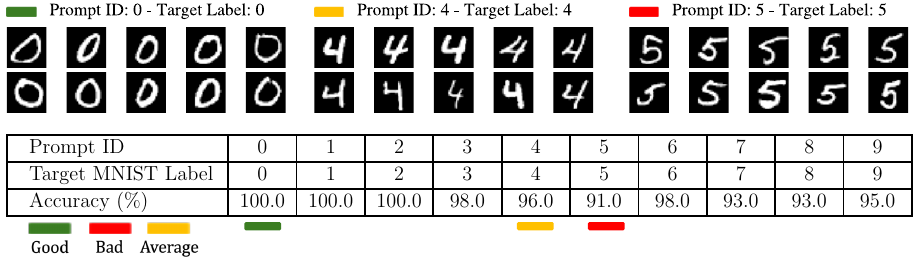}
    \caption{Performance and generated samples for three prompts from Test 1 (LL) testing conditional generation with targeted MNIST labels.}
    \label{fig:test1LL_samples}
\end{figure}

\begin{figure}
    \centering
    \includegraphics[width=.7\linewidth]{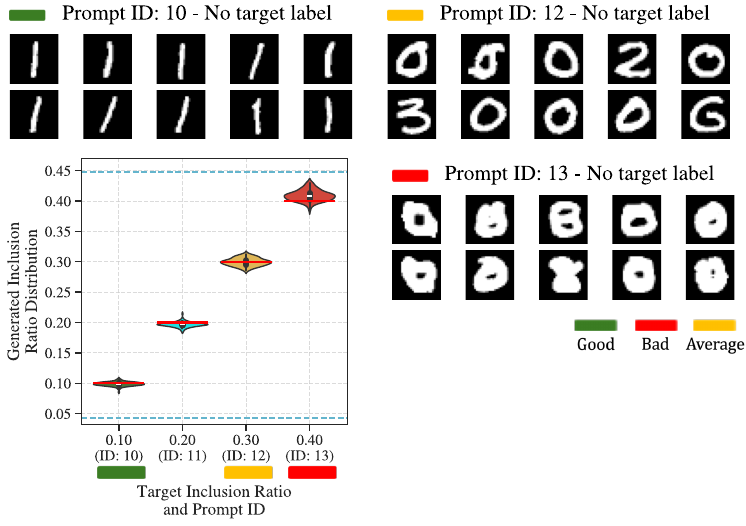}
    \caption{Performance and generated samples for three prompts from Test 2 (II) testing conditional generation with targeted inclusion ratios.}
    \label{fig:test2II_samples}
\end{figure}

\begin{figure}
    \centering
    \includegraphics[width=.85\linewidth]{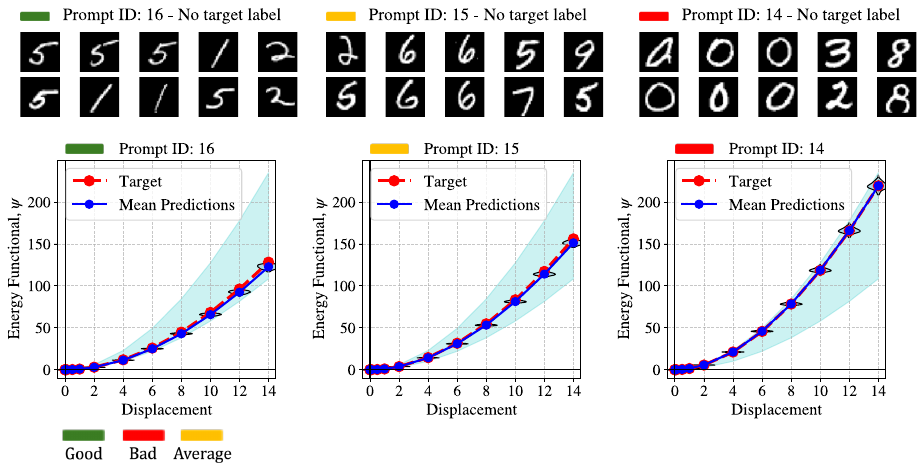}
    \caption{Performance and generated samples for three prompts from Test 3 (EE) testing conditional generation with targeted energy functionals under tension.}
    \label{fig:test3EE_samples}
\end{figure}

In Test 1 (LL), which only measures label accuracies, the model performs well across the majority of the nine prompts, demonstrating its ability to generate microstructures with target MNIST labels with over 91\% average accuracy as shown in Fig.~\ref{fig:test1LL_samples}.
In Test 2 (II), which focuses on inclusion ratios, the model also performs well, with a slightly higher MSE for prompts with higher inclusion ratios, which were less represented in the training dataset as shown in Fig.~\ref{fig:test2II_samples}.
Test 3 (EE), evaluating energy functionals, shows that the model performs rather well overall with Prompt 14 (high energy target) demonstrating the highest MSE of the three since it thee one closest to the training data range as shown in Fig.~\ref{fig:test3EE_samples}.  

For these tests, in the respective figures, we demonstrate the three selected prompts: one with the best score, one with the lowest score, and one with an average score. When there is a tie in prompt average scores, such as in Test 1 (LL), we select one randomly to demonstrate. We highlight the ranking of these prompts with labels as Good, Bad, and Average, respectively. For each prompt, we also show the first 10 samples of the 100 generated, providing a representative overview of the model's performance across different examples.

It is important to note that while the MNIST labels in Test 1 (LL) are supposed to match the target label prompts, this is not the case in Tests 2 and 3, as well as the rest of the tests that do not include L. For example, the DDPM can generate samples that have an inclusion ratio of 0.3 but appear similar to the numbers 0, 2, 3, or 6, or an interpolation between the learned digit shape distributions (Fig.~\ref{fig:test2II_samples}). This variability highlights the model's flexibility in generating diverse microstructures while still adhering to the specified inclusion ratios or energy functionals.

\begin{figure}
    \centering
    \includegraphics[width=.7\linewidth]{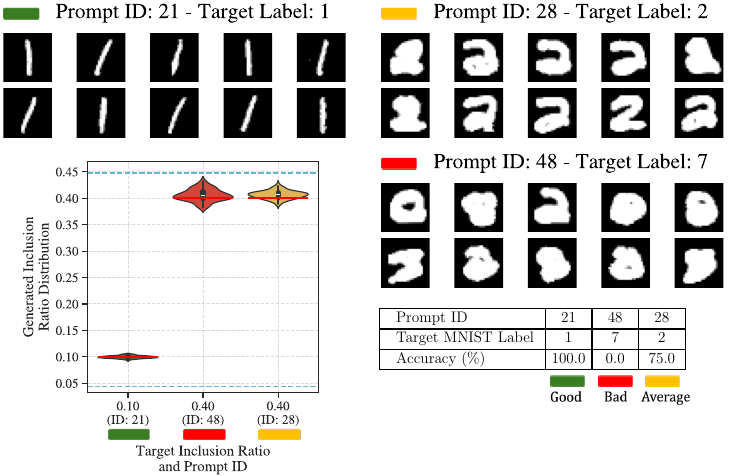}
    \caption{Performance and generated samples for three prompts from Test 4 (LI) testing conditional generation with targeted MNIST labels and inclusion ratios.}
    \label{fig:test4LI_samples}
\end{figure}

In Test 4 (LI), which combines MNIST labels and inclusion ratios, the model demonstrates its ability to handle dual contextual inputs, generating microstructures that meet both criteria. 
The results, shown in Fig.~\ref{fig:test4LI_samples}, indicate that the model performs well in generating samples with accurate MNIST labels and inclusion ratios, though the inclusion ratios slightly increase the mean squared error (MSE) compared to single-context tests. 
For this test, we highlight three selected prompts: one with the best score, one with the lowest score, and one with an average score. The samples for each prompt show the first 10 out of 100 generated, providing a representative overview.
We also observe the architecture to struggle for the first time to generate a prompt combination - Prompt 48, for which it is tasked to generate a microstructure resembling the digit 7 while having a high inclusion ratio. 
Difficulties similar to this can be expected, as such a combination of descriptors in the data may be underrepresented -- the shape "7" is thin, and it is improbable to encounter a sample with a high inclusion ratio in the dataset. 
It is noted that this does not necessarily mean that the DDPM cannot generate a microstructure with these specifications but it is more unlikely to do so and may require the generation of more samples until it can be achieved. 

\begin{figure}
    \centering
    \includegraphics[width=.85\linewidth]{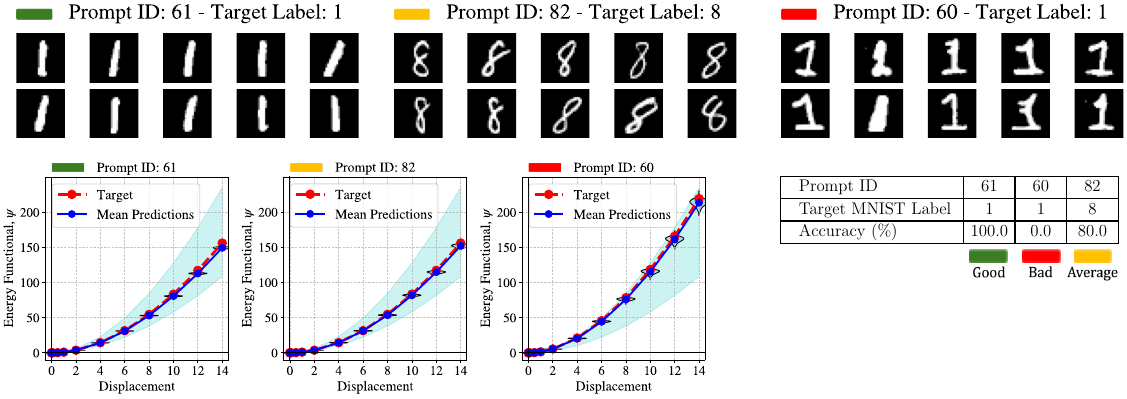}
    \caption{Performance and generated samples for three prompts from Test 5 (LE) testing conditional generation with targeted MNIST labels and energy functionals under tension.}
    \label{fig:test5LE_samples}
\end{figure}

Test 5 (LE) evaluates the model's performance in generating microstructures based on MNIST labels and energy functionals. As depicted in Fig.~\ref{fig:test5LE_samples}, the model effectively generates samples that align with both the specified MNIST labels and energy functional targets. Similar to Test 4, we demonstrate three selected prompts with varying scores, labeled as Good, Bad, and Average. 
It is also highlighted, after inspecting the generated samples manually for a few prompt cases, the DDPM may succeed in achieving the generation target but is erroneously ranked poorly by the surrogate model system.
For example, the results for Prompt 60 with an MNIST target label 1 and a high energy functional target are ranked poorly. The MSE for the energy functional is rather low, but the apparent predicted MNIST class accuracy is very poor.
However, after inspection of the samples, we identify this is not the case. 
Many of the generated Prompt 60 samples do resemble the digit 1 as it can be seen in Fig.~\ref{fig:test5LE_samples}, but they have been misclassified by the classifier surrogate model.
This is probably attributed to the DDPM learning the underlying shape distribution of the dataset better than the classifier, and is indicative of the generative capacity of the architecture to interpolate between classes of topologies.
To rectify these misclassification errors, fine-tuning of the classifier weights on generated and re-labeled data could be considered in future work.
However, it noted that in this work we will consider the MNIST labels as a secondary criterion of evaluating the generator's capacity to be further investigated in the future.

\begin{figure}
    \centering
    \includegraphics[width=.85\linewidth]{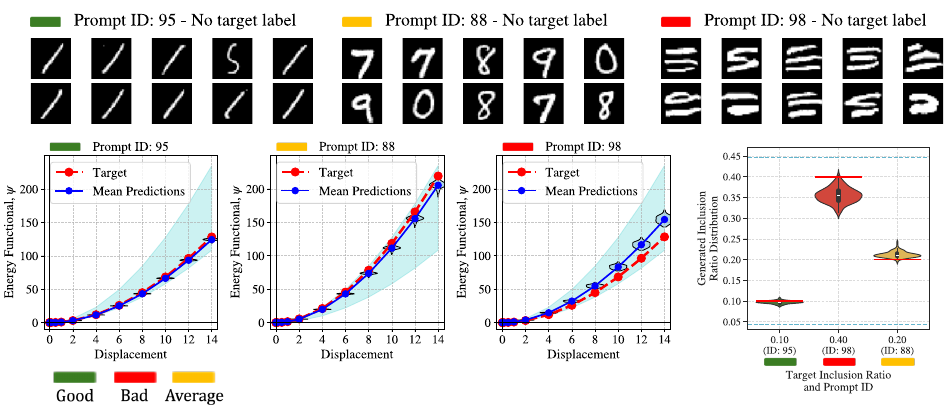}
    \caption{Performance and generated samples for three prompts from Test 6 (EI) testing conditional generation with targeted energy functionals under tension and inclusion ratios.}
    \label{fig:test6EI_samples}
\end{figure}

In Test 6 (EI), the focus is on inclusion ratios and energy functionals, testing the model's capability to generate microstructures that adhere to both these conditions simultaneously. The results in Fig.~\ref{fig:test6EI_samples} reveal that the model performs reasonably well. As with the previous tests, we display three selected prompts to illustrate the model's performance across different conditions, showing the first 10 samples generated for each prompt.
Specifically, we highlight the worst performing Prompt 98 to demonstrate when the generation targets are those of unlikely combinations (low energy functional - high inclusion ratio), the generation may fail completely, and the samples may appear to not belong to any of the underlying topology classes.

\begin{figure}
    \centering
    \includegraphics[width=.7\linewidth]{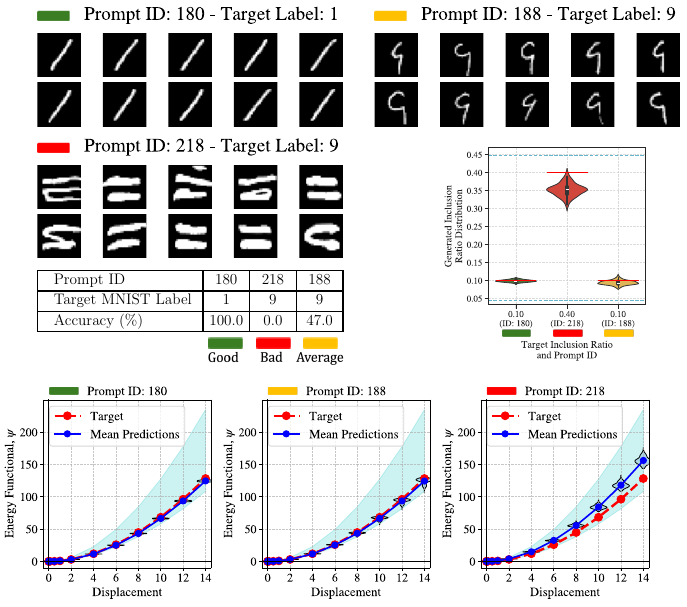}
    \caption{Performance and generated samples for three prompts from Test 7 (LEI) testing conditional generation with targeted MNIST labels, energy functionals under tension, and inclusion ratios.}
    \label{fig:test7LEI_samples}
\end{figure}

In Test 7 (LEI), which combines MNIST labels, energy functionals, and inclusion ratios, the model faces the most complex generation tasks. 
The results, shown in Fig.~\ref{fig:test7LEI_samples}, indicate a high variance in performance across the prompts. Some prompts, well-represented in the training dataset, perform exceptionally well, while others perform poorly or fail completely. 
For instance, Prompt 218 struggles to generate valid samples, whereas Prompt 180 succeeds remarkably. 
These variations can be attributed to the representation of the prompts in the training data and the inherent difficulty of combining multiple contextual factors.
For this test, we display three selected prompts with varying scores, labeled as Good, Bad, and Average, to illustrate the range of performance. 
Each figure includes the first 10 samples generated for each prompt, providing a detailed view of the model's capabilities and limitations. 
It is expected that some prompts perform poorly, either due to insufficient representation in the training dataset or because the specific combinations of inclusion ratios and energy responses under tension are physically improbable.

A more comprehensive study of the similarity between dataset descriptor distributions is outside the scope of this work and will be considered in future research. 
This future work could include mechanisms to warn users when a target microstructure combination is improbable or not physically feasible.

\FloatBarrier

\subsection{Property distribution and filtering of generated samples} 
\label{sec:filtering}
In this section, we demonstrate the effect of the distribution and the filtering of the generated samples per test. 
We show the probability densities of all generated microstructures and illustrate how the filtering process looks when applying our model filter to select samples with inclusion ratios and mean squared errors less than the target thresholds. 
Specifically, we applied a cut-off MSE of 0.01 for the inclusion ratio and 0.001 for the predicted energy functional.

We generated a large number of microstructures (21,900) of the previous example and then filtered them based on these criteria, adjusting hyperparameters to control the flexibility and accuracy of our generated microstructures. 
For each prompt, we evaluate how many microstructures were initially generated and how many remained after applying the filtering criteria. 
This process ensures that only the most accurate microstructures, according to our surrogate model system, are considered.

It is noted that the classifier may not always accurately predict the labels of the generated microstructures, and the energy functional predictions could be further validated with forward finite element simulations. 
We demonstrate the results of the filtering system using the post-processed calculated inclusion ratio and predicted energy functional under tension MSE as the criteria to filter out generated microstructures. 
We opt not to account for the results of the topology label classification due to the observed misclassifications discussed in the previous section.
As shown in our previously published work \citep{vlassis2023denoising}, the CNN surrogate model's accuracy is generally considered sufficient for assessing the performance of the generated microstructures compared to their respective FEM predicted counterparts. 
Future work may involve more stringent filtering using more robustly trained surrogate models.

\begin{figure}
    \centering
    \includegraphics[width=1\linewidth]{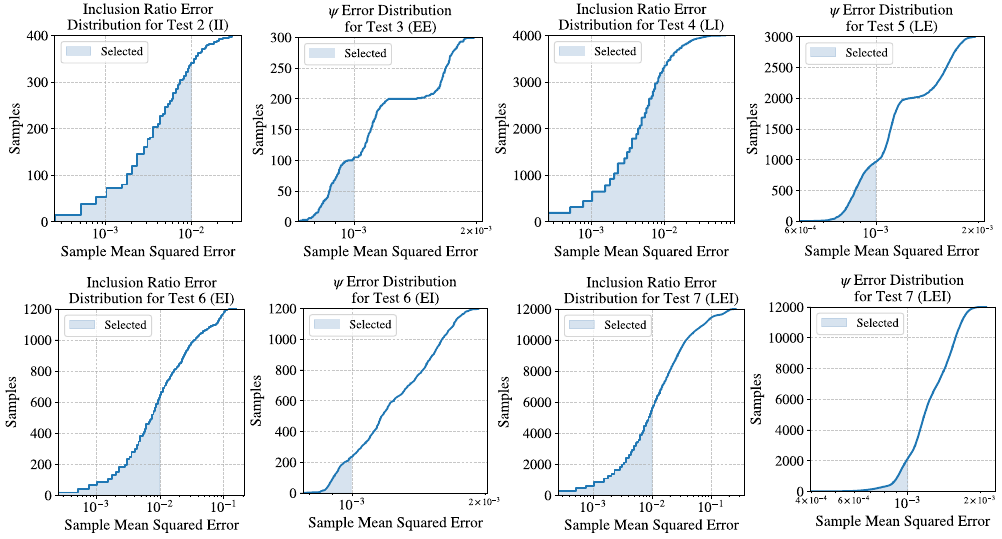}
    \caption{Empirical cumulative distribution function (eCDF) plots demonstrating the overall inclusion ratio and energy functional MSEs for Tests 2 through 7. The selected cut-off MSEs for the filtering surrogate model system are also demonstrated.}
    \label{fig:error_distributions}
\end{figure}

The filtering results of the tests involving inclusion ratio and energy functional prompts are as follows. 
In Test 2 (II), out of the 400 generated samples, 340 passed the inclusion ratio cutoff. In Test 3 (EE), from 300 generated samples, 105 passed the energy functional cutoff. 
For Test 4 (LI), out of 4000 samples, 3355 passed the inclusion ratio cutoff. 
In Test 5 (LE), out of 3000 samples, 965 passed the energy functional cutoff. 
Test 6 (EI) showed that out of 1200 generated samples, 235 passed the energy functional cutoff, 648 passed the inclusion ratio cutoff, and 213 passed both cutoffs. 
Finally, in Test 7 (LEI), out of 12000 samples, 2124 passed the energy functional cutoff, 5620 met the inclusion ratio cutoff, and 1567 samples passed both criteria. 
The empirical cumulative distribution function (eCDF) plots of the MSE distributions for these tests can be seen in Fig.~\ref{fig:error_distributions}, which demonstrates the overall inclusion ratio and energy functional MSEs for these tests, along with the selected cutoff MSEs for the filtering surrogate model system.
It is highlighted that the cut-off MSE selection is a hyperparameter that is defined based on the application at hand.
Also, filtering out a large number of generated samples does not mean that the DDPM architecture is less capable of generating samples of these specifications but that it would require relatively more sampling and filtering iterations until a satisfactory quantity of them is acquired.


\section{Ensemble test of framework} 
\label{sec:ensemble}
\begin{figure}
    \centering
    \includegraphics[width=1\linewidth]{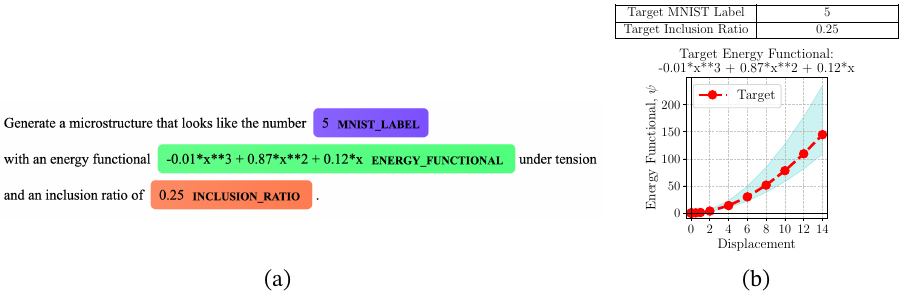}
    \caption{Steps 1 and 2 of the ensemble framework. (a) Step 1: The context named entities are identified by the NER. (b) Step 2: The context named entities are evaluated as numerical values to be then passed to the DDPM.}
    \label{fig:evaluate_ner}
\end{figure}

In this section, we demonstrate the ensemble framework that combines the NLP and DDPM components to generate targeted microstructures starting from commands in natural language. 
The sole input of the ensemble framework is a text prompt. 
We input the command "Generate a microstructure that looks like the number
5 with an energy functional $-0.01 * \mathrm{x} * * 3+0.87 * \mathrm{x} * * 2+0.12 * \mathrm{x}$ under tension and an inclusion ratio of 0.25." as shown in \ref{fig:evaluate_ner}(a). 
It is noted that this prompt is selected for demonstration purposes. 
The two components have been previously separately validated in the previous sections.

\begin{figure}
    \centering
    \includegraphics[width=1\linewidth]{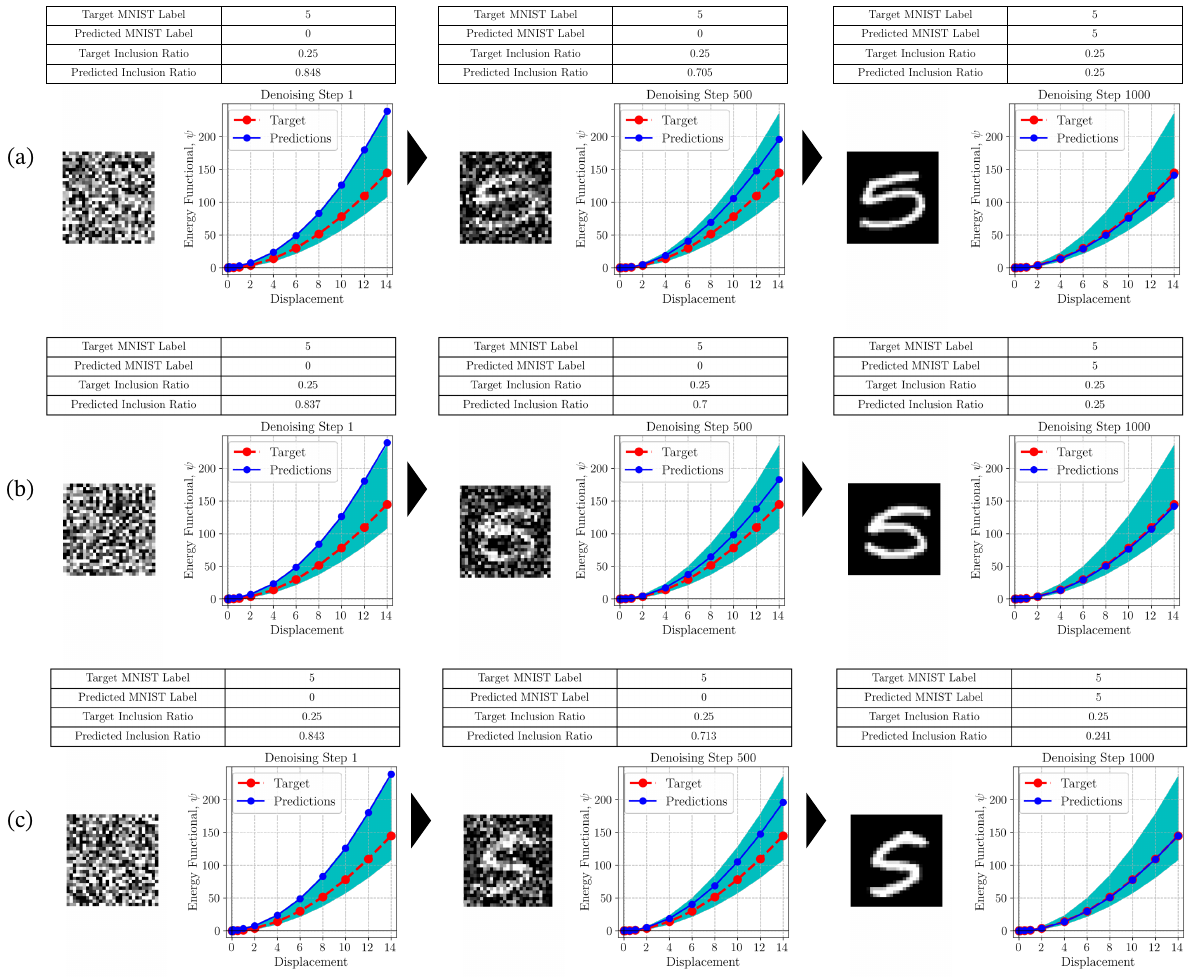}
    \caption{Steps 3 and 4 of the ensemble framework for the context prompt of Fig.~\ref{fig:evaluate_ner} for three generated samples (a, b, and c) for denoising steps 1, 500, and 1000. Step 3: Samples are generated by the DDPM. Step 4: The samples are validated by the classifier NN, the inclusion ratio calculation, and the CNN surrogate model.}
    \label{fig:converging}
\end{figure}

The prompt input is first passed through the trained NER model. 
The NER model successfully recognizes the entities MNIST label (5), inclusion ratio (0.25), and energy functional symbolic expression ($-0.01 * \mathrm{x} * * 3+0.87 * \mathrm{x} * * 2+0.12 * \mathrm{x}$). 
The extracted entities are stored as string variables. 
To acquire an acceptable input format for the DDPM, we evaluate each extracted string. 
The topology label string is evaluated as an integer, and is one-hot encoded -- turned into vector $\left[0,0,0,0,0,1,0,0,0,0\right]$.
The inclusion ratio is evaluated as a float. 
The symbolic expression is evaluated for the 12 displacement steps from the Mechanical MNIST dataset as discussed in Section~\ref{sec:database}. 
The identified and evaluated descriptors are visualized in Fig.~\ref{fig:evaluate_ner}(b). 

The evaluated variables are scaled and input to the DDPM. 
The batch size of generated samples can also be selected at this point. 
The DDPM is tasked to continuously generate microstructures until the requested number of samples with the desired properties are acquired. 
This involves the validation of the batch-generated samples with the surrogate model filter system for set cut-off mean-square errors, as defined in Section~\ref{sec:surrogate_models}.
The results of the generation of this input prompt (Fig.~\ref{fig:evaluate_ner}) are demonstrated in Fig.~\ref{fig:converging}. 
For the three generated structures, we also demonstrate the convergence of the generated descriptors over 1,000 denoising steps, starting from random noise (step 1) to generated and validated sample (step 1,000).

\FloatBarrier
\section{Conclusion}
\label{sec:conclusion}
This work presents a framework that integrates NLP and DDPM for targeted microstructure design using natural language commands. The NLP component utilizes a pretrained LLM to generate and augment text descriptors of microstructures through contextual data augmentation. A retrained NER pipeline then extracts key descriptors, such as topology labels, inclusion ratios, and energy functionals, from user-provided natural language inputs. These extracted descriptors are fed into the DDPM, which generates microstructures by reversing a diffusion process to meet the specified design targets.
The key findings demonstrate the framework's effectiveness in generating microstructures with specific properties based on natural language commands. The modular structure of the NLP and DDPM components allows for adaptation to a variety of datasets and design tasks, providing flexibility for different applications. The use of a surrogate model system further improves the output of the framework by ranking and filtering the generated microstructures to ensure they closely align with the desired properties specified by the user.

\section{Data Availability}
The computer code and data that support the findings of this study are available from the corresponding author upon request. 

\section{Credit authorship contribution statement}
Nikita Kartashov: Software, Validation, 
Formal analysis, Data curation, Writing.
Nikolaos Vlassis: Conceptualization, Methodology, Software, Validation, 
Formal analysis, Data curation, Writing.

\bibliographystyle{plainnat}
\bibliography{main}

\end{document}